\RequirePackage{fix-cm}

 \documentclass[showkeys,preprint]{revtex4}          
\usepackage{graphicx}
\usepackage{amssymb}
\usepackage{amsbsy}
\usepackage{ulem}
\usepackage{textcomp}
\usepackage{hyperref}
\usepackage[hang,small]{caption}
\usepackage{amssymb}
\usepackage{amsmath}
\usepackage{diagbox}

\usepackage{subfigure}
\usepackage{color}
\usepackage{enumitem}
\usepackage{cases}

\begin{document}

\title[]{Computation of thermal properties via 3D homogenization of multiphase materials using FFT-based accelerated scheme
}

\author{Sophie Lemaitre}
\author{Vladimir Salnikov} 
\author{Daniel Cho\"i}
\author{Philippe Karamian}

\affiliation{Nicolas Oresme Mathematics Laboratory \\ University of Caen Lower Normandy\\ 
  CS 14032, Bd. Mar\'echal Juin,  BP 5186\\
  14032, Caen Cedex,  France \\
  \email{vladimir.salnikov@unicaen.fr, daniel.choi@unicaen.fr, philippe.karamian@unicaen.fr}
}



\begin{abstract}
In this paper we study the thermal effective behaviour for 3D multiphase composite material consisting of three isotropic phases which are the matrix, the inclusions and the coating media. For this purpose we use an accelerated FFT-based scheme initially proposed in Eyre and Milton (1999) to evaluate the thermal conductivity tensor.
Matrix and spherical inclusions media are polymers with similar properties whereas the coating medium is metallic hence better conducting. Thus, the contrast between the coating and the others media is very large.
For our study, we use RVEs (Representative volume elements) generated by RSA (Random Sequential Adsorption) method developed in our previous works, then, we compute effective thermal properties using an FFT-based homogenization technique validated by comparison with the direct finite elements method. We study the thermal behaviour of the 3D-multiphase composite material and we show what features should be taken into account to make the computational approach efficient.


\end{abstract}

\maketitle

\section{Introduction and motivations}

In this paper, we study the thermal conductivity tensor using the 3D homogenization method for materials constituted of 3 phases.
Since composite materials are widely used in modern industry, there has been a number of works 
studying their effective properties: they concern descriptions of modelling approaches as well as results 
of experimental measurements. The purpose of this paper is to explore the range of application and features
for existing modelling approaches, and at the same time the work is motivated by 
precise application within the framework of an industrial project. \\
The importance of modelling for the analysis of composite materials results from
various challenges in carrying out experimental works: such problems include 
purely technical difficulties as well as financial optimization of the applied activities.
It is thus necessary to master modelling approaches that are reliable, comparable with experiment, 
and still affordable from the point of view 
of computational efficiency.
 Our strategy in this paper is related to the idea of stochastic homogenization. 
 We consider samples of a composite material containing coated inclusions;  
 and in order to take into account possible imperfections and random factors, 
 we average the result for a series of tests representing the same macro characteristics. 
 For this work we have chosen to study relatively simple geometry of inclusions: spherical shape with uniform coating. On  the one hand this permits to concentrate on the details of 
 application of the adopted modelling technique, on the other hand this can still provide some 
 insight on the influence of morphology on the effective properties of composite materials.\\
The paper is organized as follows. In the next section, we explain the RSA (Random Sequential Adsorption) method chosen for the generation of RVEs (Representative volume elements) proposed in \cite{Salnikov1} and we give the context of the study. Next, we recall the model used with the computational method. In the third section we compare the FFT (Fast Fourier Transform)-based homogenization technique and the finite elements method (FEM) for simple geometries and we motivate our choice of the FFT-based `accelerated scheme'.
The last section describes the numerical results. 
We conclude by recapitulating the main points of this work and describing the work in progress.


\section{Sample generation and setting of the study}
\begin{figure}[!htbp]
\begin{center}
\includegraphics[scale=0.7]{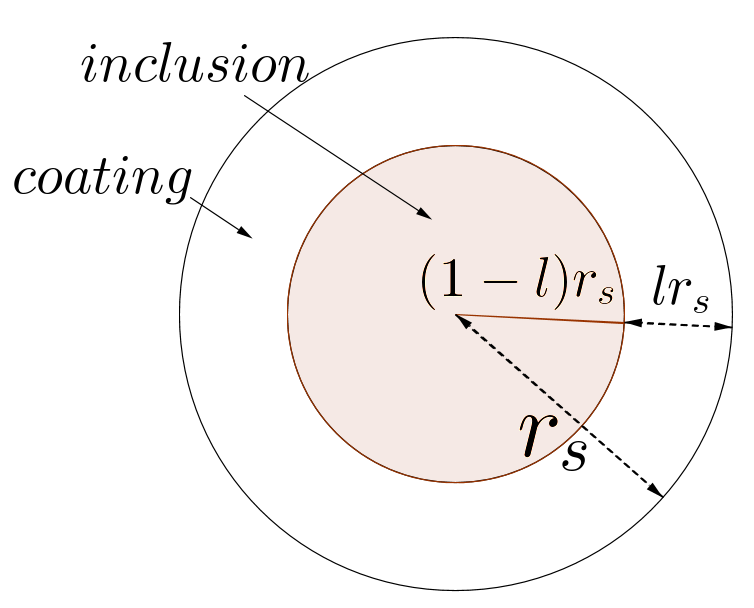}
\caption{Sphere made of a spherical inclusion with a coating parametrized by a layer $l$}
\label{fig1}
\end{center}
\end{figure}

\begin{figure}[!htbp]
\begin{center}
\includegraphics[height=0.15\textheight]{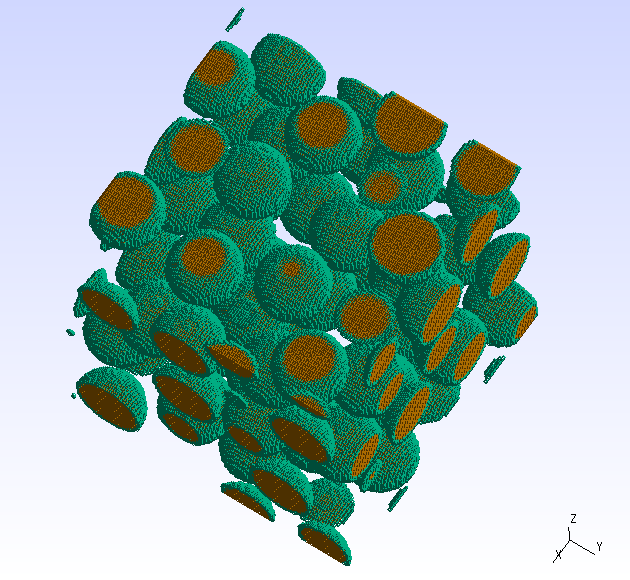}
\hspace{0.3cm}
\includegraphics[height=0.15\textheight]{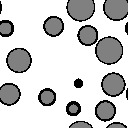}
\hspace{0.3cm}
\includegraphics[height=0.15\textheight]{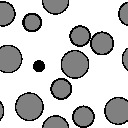}
\caption{Example of one RVE meshed with 3 phases generated by RSA method and two sections in $128^3$ voxels representing the top face and one of these center sections}
\label{figRVE}
\end{center}
\end{figure}

In this section, we describe the method used to generate RVEs with three phases. As we have outlined in the introduction, we consider a material with the matrix, the spherical inclusions and the coating media.\\ 
We proceed in two steps. First, we generate RVEs using the tool developed by V. Salnikov et al. in \cite{Salnikov1}. With this tool using the RSA algorithm  generation method, we are able to generate three phased RVEs containing up to $100$ identical spherical inclusions and up to $40\%$ in volume fraction. In the cited paper, it has been shown that the method is very efficient being able to generate such RVEs in fractions of a second. The described algorithm produces a list of inclusions in a "vector" form namely for each sphere coordinates of the center (in 3D) and the radius. The spheres do not overlap. In the second step, we use a voxelization tool (\cite{CSMA}) to obtain a voxelized sample with three phases. We add a coating defined by the user for each spherical inclusion. The figure \ref{fig1} shows a scheme representing a section of a sphere with the notations used throughout the article. In the sequel, we note $l$ the layer which is the ratio between the coating width and the radius of a sphere  $r_{s}$, the layer is thus a number between $0$ and $1$.
On the figures \ref{figRVE}, we see one 3D RVE voxelization and two of theses sections.

\noindent The setting of the study is the following: we fix the parameters like the volume fraction of spheres $f_{sp}$ equal to $30\%$ (which takes into account spherical inclusion and coating) and the phases conductivity tensor for each phase equals to $kI$ where $k$ is a positive number (it takes 3 different values for the 3 materials and $I$ stands for the identity matrix in dimension 3. \\
In this setting, we study how the thermal properties of a composite vary depending on a geometrical parameter denominated as the layer keeping in mind that the spherical volume fraction is fixed here at $30\%$. For this purpose, we generated samples with different quantity of spheres and different resolutions to observe the behaviour of the composite.


\section{Computational techniques}

Consider a representative volume element $V$ with periodic boundary conditions, and denote $\theta(\boldsymbol{x})$ the temperature, $\boldsymbol{\phi}(\boldsymbol{x})$ the heat flux and $L(\boldsymbol{x})$ the thermal conductivity at any point $\boldsymbol{x}  \in  V$. \\
Fourier's law in the linear case states that: $\boldsymbol{\phi}(\boldsymbol{x})=-L(\boldsymbol{x})\: \nabla \theta(\boldsymbol{x})$. Moreover, we suppose that we are in stationary phase and without heat source so the heat flux verifies $div(\boldsymbol{\phi}(\boldsymbol{x})) = 0$. Notice that for a composite material, the thermal conductivity tensor depends on the point $\boldsymbol{x}$: the dependence is governed by microscopic geometry of the sample, namely which phase (matrix, inclusion or coating) the point belongs to. \\
The volume element $V$ is subjected to an average temperature gradient $\langle  \nabla \theta(\boldsymbol{x})\rangle=\nabla\Theta$ which induces local temperature gradient $\nabla \theta(\boldsymbol{x})$ and heat flux $\boldsymbol{\phi}(\boldsymbol{x})$  inside the RVE. The effective constitutive relations of the composite are the relations between the (spatial) average flux $\Phi$ and average temperature gradient $\nabla \Theta$.\\

\noindent It is a common practice to introduce a constant reference tensor $L^{0}$ to solve the problem of recovering the local fields and gradients.
Introducing this homogeneous linear tensor, we can rewrite the problem as:
\begin{equation}
\left \{
\begin{array}{l}
\boldsymbol{\phi}(\boldsymbol{x})=-L^{0}(\boldsymbol{x}) \nabla \theta(\boldsymbol{x}) + \boldsymbol{\tau}(\boldsymbol{x}), ~ \forall \boldsymbol{x} \in V \\
div(\boldsymbol{\phi}(\boldsymbol{x})) = 0, ~ \forall \boldsymbol{x} \in V \\
\boldsymbol{\tau}(\boldsymbol{x})= -(L(\boldsymbol{x})-L^{0})\nabla \theta(\boldsymbol{x}), ~ \forall \boldsymbol{x} \in V 
\end{array}
\right.
\end{equation}
where $\boldsymbol{\tau}$ denotes the polarization tensor.\\
The solution of (1) can be expressed in real and Fourier spaces, where $\widehat{\bullet}$ denotes the Fourier image of $\bullet$ and $\xi_{j}$'s are the coordinates in the Fourier space, respectively, with the help of the periodic Green operator $\Gamma^{0}$ associated with $L^{0}$ we can also write:
\begin{equation}
\nabla \theta(\boldsymbol{x})= \Gamma^{0}(\boldsymbol{x})\nabla \boldsymbol{\tau}(\boldsymbol{x}) ~~~ \forall \boldsymbol{x} \in V 
\end{equation}
or in Fourier space:
\begin{equation}
\widehat{\nabla \theta}(\boldsymbol{\xi})= \widehat{\Gamma}^{0}(\boldsymbol{\xi})\widehat{\boldsymbol{\tau}} (\boldsymbol{\xi})~~~ \forall \boldsymbol{\xi} \neq 0 , ~ \widehat{\nabla \theta}(\boldsymbol{0})=0
\end{equation}
The equations (2) and (3) give the Lippmann-Schwinger integral equations  in real and Fourier spaces respectively.
The Green operator $\widehat{\Gamma}^{0}$ is easily expressed and computed in the Fourier space by:
$$
\widehat{\Gamma}^{0}_{ij}(\boldsymbol{\xi})=\dfrac{\boldsymbol{\xi}_{i}\boldsymbol{\xi}_{j}}{\sum\limits_{m,n}L^{0}_{mn}\boldsymbol{\xi}_{m}\boldsymbol{\xi}_{n}}
$$

\noindent The Lippmann-Schwinger equation can be solved iteratively using the algorithm based on the accelerated scheme proposed by Eyre and Milton in \cite{Eyre and Milton} and written as algorithm 3 in \cite{Salnikov2} for the elastic case. For the thermal conductivity case, we obtain with adapted notations the following algorithm:\\
\newpage
\noindent\textbf{Algorithm:} \textit{FFT-based homogenization scheme for thermal conductivity case}\\
\noindent\textit{Initialize $\nabla \theta^{0}(\boldsymbol{x})\equiv \nabla \Theta$}\textit{, fix the convergence criterion acc.}\\
\textit{while (not converged)}
      \newline \begin{tabular}{c|l}   & \parbox{0.98\linewidth}{
      \begin{enumerate}
       \item \textit{Convergence test:}\\
       \textit{if (}$\epsilon_{comp} < acc)$ \textit{compute} $\boldsymbol{\phi}^{n}(\boldsymbol{x})=- L(\boldsymbol{x})\nabla \theta^{n}(\boldsymbol{x})$, $\widehat{\boldsymbol{\phi}}^{n}= FFT(\boldsymbol{\phi}^{n})$,\\
       $\epsilon_{eq}=\sqrt{\langle\Vert\boldsymbol{\xi} \widehat{\boldsymbol{\phi}}^{n}(\boldsymbol{\xi})\Vert^{2}\rangle}/\Vert\widehat{\boldsymbol{\phi}}(\boldsymbol{0})\Vert$\\
       \textit{if} $(\epsilon_{eq} <acc)$ \textit{then converged, stop}
       \item $\boldsymbol{\tau}^{n}(\boldsymbol{x})= -(L(\boldsymbol{x}) + L^{0})\nabla \theta(\boldsymbol{x})$
       \item $\widehat{\boldsymbol{\tau}}^{n}= FFT(\boldsymbol{\tau}^{n})$
       \item $\widehat{\nabla \theta}^{n}_{comp}(\boldsymbol{\xi})=-\widehat{\Gamma}^{0}(\boldsymbol{\xi})\widehat{\boldsymbol{\tau}}^{n}(\boldsymbol{\xi}), \boldsymbol{\xi} \neq 0$,  $\widehat{\nabla \theta}^{n}_{comp}(\boldsymbol{0})=\nabla \Theta$
       \item $\nabla \theta^{n}_{comp}= FFT^{-1}(\widehat{\nabla \theta}^{n}_{comp})$
       \item $\epsilon_{comp}=\sqrt{\langle\Vert \nabla \theta^{n}-\nabla \theta^{n}_{comp}\Vert^{2}\rangle} /\Vert \nabla \Theta\Vert$
       \item $\nabla \theta^{n+1}(\boldsymbol{x})= \nabla \theta^{n}(\boldsymbol{x}) - 2(L(\boldsymbol{x})-L^{0})^{-1}L^{0}(\nabla \theta^{n}_{comp}(\boldsymbol{x})-\nabla \theta^{n}(\boldsymbol{x}))$
       \end{enumerate}
        }\end{tabular}

\vspace{0.3cm}    
\noindent When this algorithm converges, we compute $\langle \boldsymbol{\phi}(\boldsymbol{x})\rangle$ and we obtain $L_{hom}$ using the following equality: $\langle \boldsymbol{\phi}(\boldsymbol{x})\rangle =-L_{hom}\langle \nabla \theta(\boldsymbol{x})\rangle=-L_{hom} \nabla  \Theta $.\\
There are several FFT-based numerical schemes like the 'basic scheme', the 'dual scheme', the augmented Lagrangian scheme, the 'polarization scheme' and the 'accelerated scheme'. We have chosen the last one for its computational efficiency like in the elastic case \cite{Salnikov2}. We justify this choice by relying on Moulinec and Silva in \cite{Moulinec and Silva}; they have shown that the accelerated scheme is the optimal compromise except for composite with infinite contrast. In this paper, the contrast of the materials studied is between $5$ and  $2.10^{3}$.

\section{Comparison FEM and FFT method}

\begin{figure}[!htbp]
\begin{center}
\includegraphics[height=0.2\textheight]{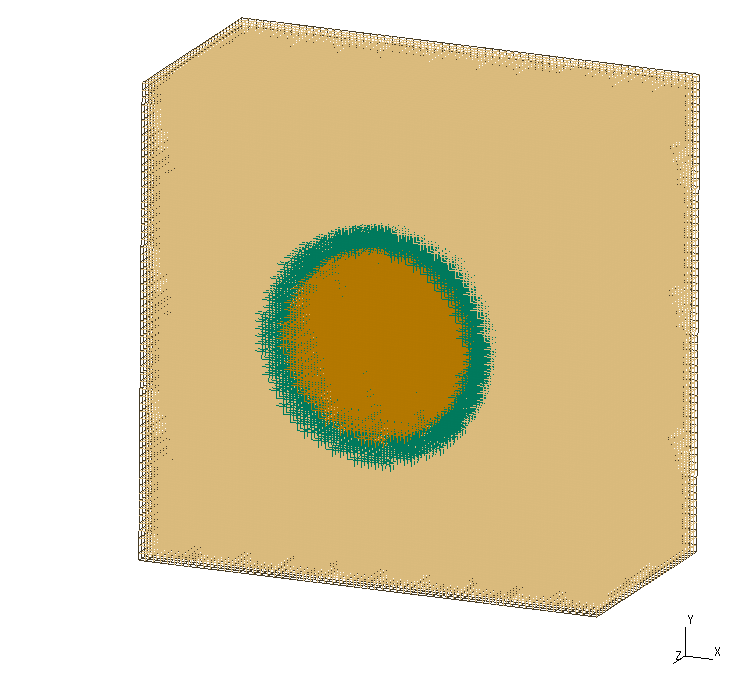}
\hspace{0.3cm}
\includegraphics[height=0.2\textheight]{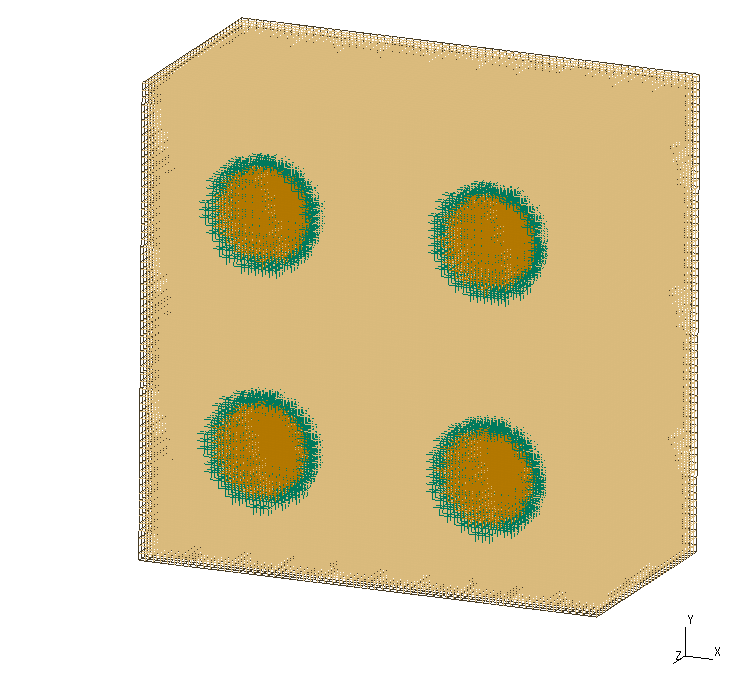}
\caption{Two cuts representing RVE with one coated sphere and four coated spheres in resolution $64^3$ voxels with sufficient coating}
\label{figSphere}
\end{center}
\end{figure}

We briefly describe how we evaluate the homogenized thermal properties of the composite using the FEM. We use a double-scale method described by Sanchez-Palencia \cite{Sanchez} and Bensoussan, Lions and Papanicolaou \cite{Bensoussan}. It deals with setting a multi-scale problem via an asymptotic expansion of the equations describing the behaviour of the material. In our case, we consider a double-scale problem in the context of the thermal conductivity. The composite material is associated to the macroscopic scale and the RVE to the microscopic one. We impose periodic boundary conditions.\\
In order to compare both approaches, we study simple cases namely: one sphere or four spheres centred as it is shown on figure \ref{figSphere}.

\begin{figure}[htbp]
\begin{center}
\includegraphics[height=0.2\textheight]{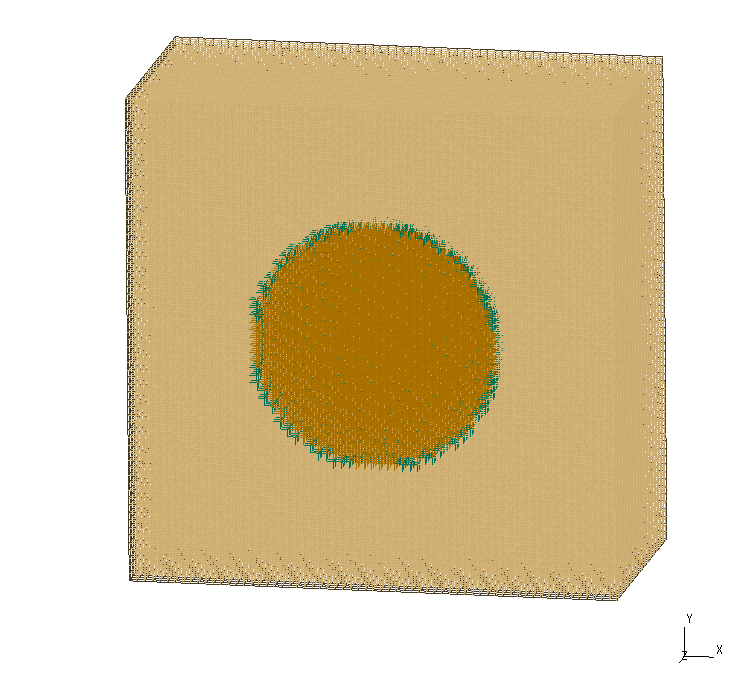}
\caption{A cut representing RVE with one sphere with thin coating}
\label{figThinC}
\end{center}
\end{figure}

\noindent We generate RVEs in resolution $64^3$ and $128^3$ voxels and with several thickness of the layer. For each generated RVE, we construct a mesh
based on the voxels then each voxel defines a cub8 element and we use a python code to compute the thermal homogenized tensor.\\
The RVEs with a thin coating in low resolution namely $64^3$ voxels or lower, exhibits some voids on the coating as one can notice on figure \ref{figThinC}. In these cases, the results obtained by the two methods show some deviation which can reach $20\%$.

\noindent However, for the RVEs with sufficient coating such in figure \ref{figSphere}, the results obtained by the different methods are very similar, the discrepancy between both methods does not exceed $1\%$. An important remark is that with more complex geometries, the 3D homogenization requires a lot of memory resources and takes a lot of CPU time with the FEM whereas the FFT method has a lower cost in time and it allows the computation in $256^{3}$ voxels and even in $512^{3}$ voxels (using a computation cluster). Thus we extend the FFT method to more complex geometries namely with at least $30$ spheres located randomly without intersection.


\section{Numerical simulations}

Numerical computations are made for 3 resolutions: $64^{3}$, $128^{3}$ and $256^{3}$ voxels. According to these resolutions, for the same geometries, the number of voxels representing each phase increases. We also take $30$ up to $100$ spheres for $n_{sp}$.
We recall that the layer is a real number between $0$ and $1$. We make the tests with $l$ equal to $0.02$ up to $0.08$ and to $0.1$ up to $0.9$. In fact, from a layer equal to $0.2$, the coating reaches almost the half of the sphere volume as we see in the table \ref{tab0}.  The coating volume fraction is evaluated by the equality : $f_{coating}=(1-(1-l)^3)\times0.3$, where $0.3$ denotes the spheres volume fraction.

\begin{center}
\begin{table}
\begin{tabular}{|c|c|}
\hline 
layer &  $f_{coating}$ \\ \hline 
0.02 & 0.0176 \\  
0.04 & 0.0346 \\
0.06 & 0.0508 \\
0.08 & 0.0664\\
0.1 &  0.0813\\
\hline \hline
0.2 &  0.1464\\
0.3 & 0.1971\\
0.4 &  0.2352\\
0.5 &  0.2625\\
0.6 & 0.2808\\
0.7 &  0.2919\\
0.8 &  0.2976\\
0.9 & 0.2997\\
  \hline
\end{tabular}
\caption{Coating volume fraction depending on the thickness of the layer}
\label{tab0}
\end{table}
\end{center}

\noindent For each set of ($Res$, $n_{sp}$, $l$), we generate 10 different samples in order to make the average of obtained macroscopic conductivity tensor.

\noindent To compute the thermal conductivity tensor, we recall that the materials are supposed to be isotropic namely with a tensor equal to $kI$, where $k$ is a positive number and $I$ stands for the identity matrix in dimension 3. 
We take: 
 $$L_{matrix \:phase}= I \:,\:L_{inclusion\: phase}= 0.2 I \:,\:
L_{coating \:phase}= 400 I$$
With theses values, the higher contrast is $2.10^{3}$ and the lower is $5$.
\noindent We deliberately choose these coefficients for industrial applications with matrix and inclusion phases poorly conducting and coating phase highly conducting.
According to \cite{Eyre and Milton} and \cite{Moulinec and Sucquet 2001} the constant reference thermal tensor $L^{0}$ is set to:
 $$L^{0}=-\sqrt{\underset{x \in \{1,0.2,400\}}{min(x)} \times\underset{x \in \{1,0.2,400\}}{max(x)} } \: I=-4\sqrt5 \:I$$ 

\noindent We remark that the computed thermal homogenized tensor is close to an isotropic material tensor, so we take as the homogenized thermal conductivity the mean of the diagonal elements of the tensor for the series of the 10 geometries generated as discussed above.

\noindent The curves below on figures \ref{fig.30layer}, \ref{fig.70layer}, \ref{fig.100layer} give these homogenized values.
First, we naturally observe that the homogenized thermal conductivity increases with the thickness of the layer and for all the three resolutions tested.
We can also note that the larger the resolution is, the less the dispersion is. We conclude that resolution 64 is not accurate enough.

\begin{figure}[!htbp]
\begin{center}
 \includegraphics[width=0.75\textwidth ,trim=10 0 0 0,clip=true]{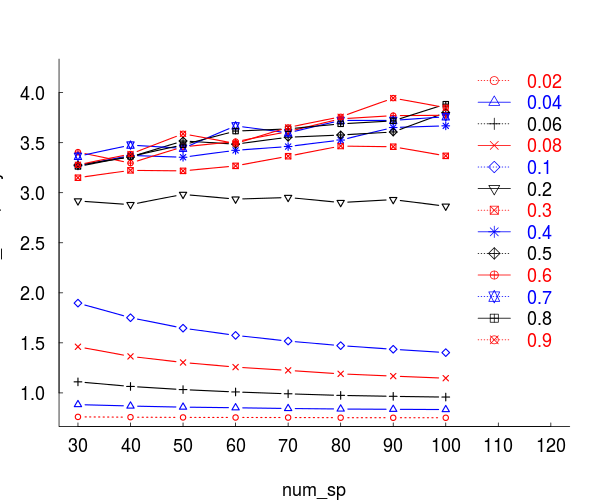}
\caption{Resolution $64^{3}$: homogenized thermal conductivity evolution vs layer for $n_{sp}$ from 30 up to 100}
\label{fig.30frac}
\end{center}
\end{figure}
\begin{figure}[!htbp]
\begin{center}
\includegraphics[width=0.75\textwidth ,trim=10 0 0 0,clip=true]{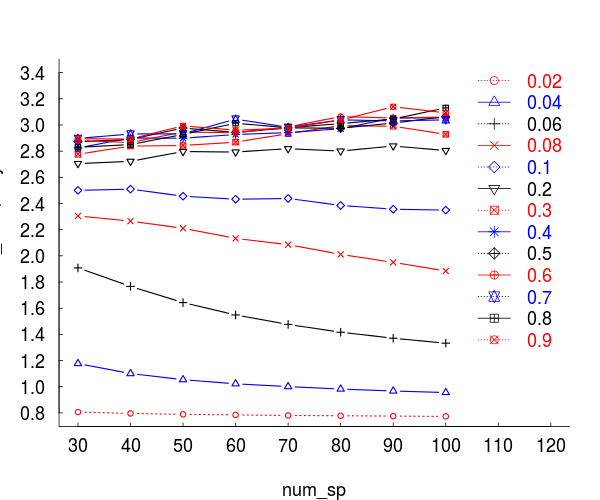}
\caption{Resolution $128^{3}$: homogenized thermal conductivity evolution vs layer for $n_{sp}$ from 30 up to 100}
\label{fig.70frac}
\end{center}
\end{figure}
\begin{figure}[!htbp]
\begin{center}
\includegraphics[width=0.75\textwidth ,trim=10 0 0 0,clip=true]{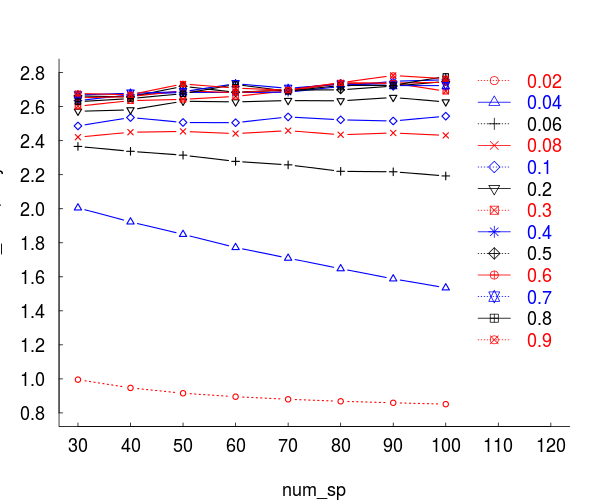}
\caption{Resolution $256^{3}$: homogenized thermal conductivity evolution vs layer for $n_{sp}$ from 30 up to 100}
\label{fig.100frac}
\end{center}
\end{figure}

\noindent We observe for very thin layers that the homogenized thermal conductivity decreases when the $n_{sp}$ increases because the coating is thin and the number of voxels representing it decrease and can be less than one pixel, i.e. it can be missed: this is shown in the tables \ref{tab1}, \ref{tab2}, \ref{tab3} where we use the equality:\\
Number of voxels = $ l \times Res \times\left( \frac{3\times 0.3\times n_{sp}}{4 \times \pi} \right)^{1/3}$, with $Res \:\in\:\{64, 128, 256\}$.\\ The values in the tables are deliberately not rounded up.

\begin{center}
\begin{table}
\begin{tabular}{|c|c|c|c|c|c|c|c|c|}
\hline
\backslashbox{$l$}{$n_{sp}$} & 30& 40 & 50 & 60 & 70 & 80 & 90& 100\\ \hline \hline 
0.02 & 0.17 & 0.16  & 0.14 & 0.14 & 0.13 & 0.12 & 0.12 & 0.11 \\  
0.04 & 0.34 & 0.31  & 0.29 & 0.27 & 0.26 & 0.25 & 0.24 & 0.23 \\
0.06 & 0.51 & 0.47  & 0.43 & 0.41 & 0.39 & 0.37 & 0.36 & 0.34 \\
0.08 & 0.68 & 0.62  & 0.58 & 0.54 & 0.52 & 0.49 & 0.47 & 0.46 \\
0.1 & 0.86 & 0.78  & 0.72 & 0.68 & 0.64 & 0.62 & 0.59 & 0.57 \\
\hline \hline
0.2 & 1.71 & 1.55  & 1.44 & 1.36 & 1.29 & 1.23 & 1.19 & 1.15 \\
0.3 & 2.57 & 2.33  & 2.16 & 2.04 & 1.93 & 1.85 & 1.78 & 1.72 \\
0.4 & 3.42 & 3.11  & 2.89 & 2.72 & 2.58 & 2.47 & 2.37 & 2.29 \\
  \hline
\end{tabular}
\caption{Voxels coating thickness for the resolution $64^{3}$ voxels}
\label{tab1}
\end{table}
\end{center}
\vspace*{0.5cm}

\begin{center}
\begin{table}
\begin{tabular}{|c|c|c|c|c|c|c|c|c|}
\hline
\backslashbox{$l$}{$n_{sp}$} & 30& 40 & 50 & 60 & 70 & 80 & 90& 100\\ \hline \hline 
0.02 & 0.34 & 0.31  & 0.29 & 0.27 & 0.26 & 0.25 & 0.24 & 0.23 \\  
0.04 & 0.68 & 0.62  & 0.58 & 0.54 & 0.52 & 0.49 & 0.47 & 0.46 \\
0.06 & 1.03 & 0.93  & 0.87 & 0.81 & 0.77 & 0.74 & 0.71 & 0.69 \\
0.08 & 1.37 & 1.24  & 1.15 & 1.09 & 1.03 & 0.99 & 0.95 & 0.92 \\
\hline \hline
0.1 & 1.71 & 1.55  & 1.44 & 1.36 & 1.29 & 1.23 & 1.19 & 1.15 \\
0.2 & 3.42 & 3.11  & 2.89 & 2.72 & 2.58 & 2.47 & 2.37 & 2.29 \\
0.3 & 5.13 & 4.66  & 4.33 & 4.07 & 3.87 & 3.70 & 3.56 & 3.44 \\
  \hline
\end{tabular}
\caption{Voxels coating thickness for the resolution $128^{3}$ voxels}
\label{tab2}
\end{table}
\end{center}

\vspace*{0.5cm}
\begin{center}
\begin{table}
\begin{tabular}{|c|c|c|c|c|c|c|c|c|}
\hline
\backslashbox{$l$}{$n_{sp}$} & 30& 40 & 50 & 60 & 70 & 80 & 90& 100\\ \hline \hline 
0.02 & 0.68 & 0.62  & 0.58 & 0.54 & 0.52 & 0.49 & 0.47 & 0.46 \\  
0.04 & 1.37 & 1.24  & 1.15 & 1.09 & 1.03 & 0.99 & 0.95 & 0.92 \\
\hline \hline
0.06 & 2.05 & 1.87  & 1.73 & 1.63 & 1.55 & 1.48 & 1.42 & 1.37 \\
0.08 & 2.74 & 2.49  & 2.31 & 2.17 & 2.06 & 1.97 & 1.90 & 1.83 \\
0.1 & 3.42 & 3.11  & 2.89 & 2.72 & 2.58 & 2.47 & 2.37 & 2.29 \\
  \hline
\end{tabular}
\caption{Voxels coating thickness for the resolution $256^{3}$ voxels}
\label{tab3}
\end{table}
\end{center}

\noindent We also note for resolution $256^{3}$ voxels that the number of spheres almost does not influence the computed values, which agrees with the definition of acceptable RVE size and means we no longer have the 'artefact' due to the voxelization. Moreover, for a layer $l$ equal to $0.08$, there is a kind of stagnation, it is shown on figures \ref{fig.30layer}, \ref{fig.70layer}, \ref{fig.100layer}. On these figures, we clearly observe two behaviours, one before the stagnation and the other one during the stagnation. This effect is less visible in resolution $128$ since it is difficult to capture the coating when the layer is too low which appears in figures \ref{voids}. From the point of view of applications we can say that it is not necessary to have a big coating to increase the homogenized thermal conductivity. There is a limit due to the sphere volume fraction and due to the type of spherical inclusions.

\begin{figure}[!htbp]
\begin{center}
 \includegraphics[width=0.9\textwidth]{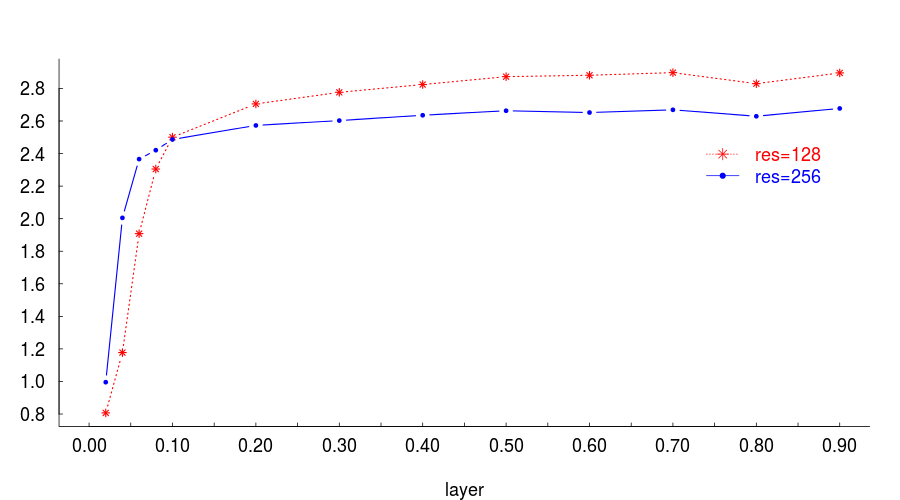}
\caption{Homogenized thermal conductivity evolution vs coating volume fraction for 30 spheres}
\label{fig.30layer}
\end{center}
\end{figure}
\begin{figure}[!htbp]
\begin{center}
\includegraphics[width=0.9\textwidth]{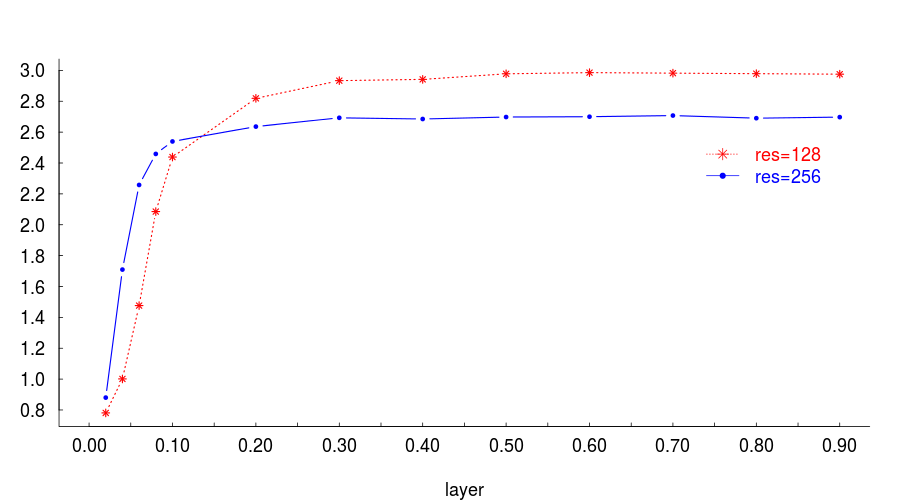}
\caption{Homogenized thermal conductivity evolution vs coating volume fraction for 70 spheres}
\label{fig.70layer}
\end{center}
\end{figure}

\begin{figure}[!htbp]
\begin{center}
\includegraphics[width=0.9\textwidth]{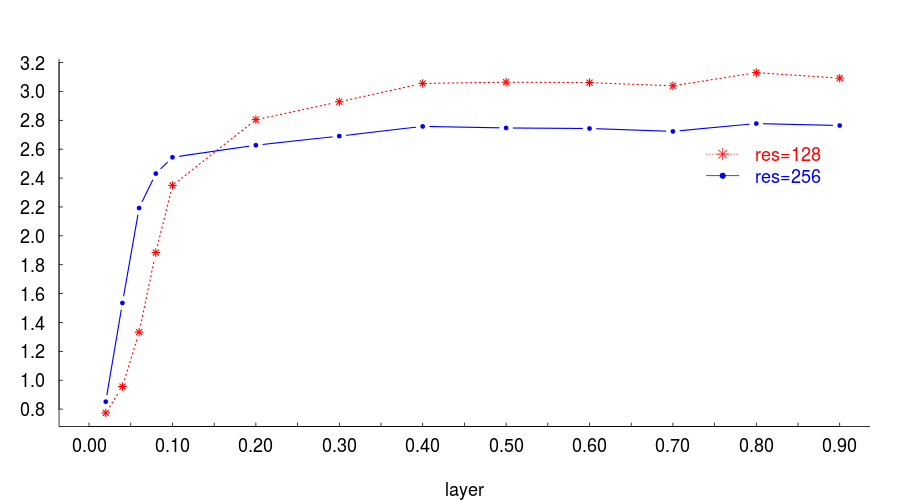}
\caption{Homogenized thermal conductivity evolution vs coating volume fraction for 100 spheres}
\label{fig.100layer}
\end{center}
\end{figure}

\begin{figure}[!htbp]
\begin{center}
\includegraphics[height=0.25\textheight]{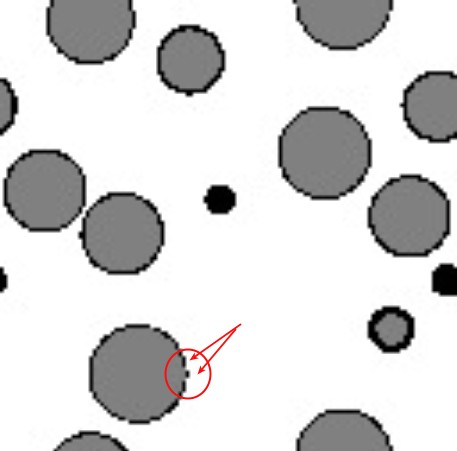}
\hspace{0.3 cm}
\includegraphics[height=0.25\textheight]{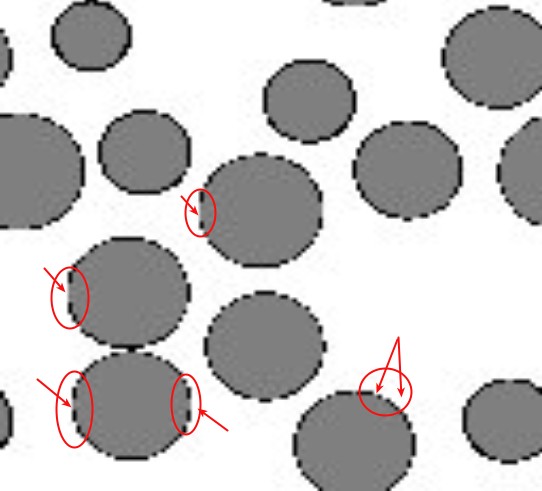}
\caption{Example of sections with lack of coating when the layer is too low}
\label{voids}
\end{center}
\end{figure}



\section{Conclusions and outlook}

To conclude, we can summarize the main ideas of this paper. We have shown that the FFT-based iterative accelerated scheme is a good tool for computing thermal effective properties of a composite generated randomly in 3D. We have restricted our study to RVE composed with only spheres but it can be made with cylinders or both spheres and cylinders. In \cite{Salnikov1}, RVEs generation algorithms in MD (Molecular dynamics) or RSA methods have been presented using both spheres and cylinders and in \cite{Salnikov2} these RVEs have been used for calculations in elastic case.\\
 We have observed that the resolution $256^{3}$ with $50-70$ spheres and an sufficient thickness of the layer gives good results. We have also noted that the coating must be thick enough in order to be "captured". In our work in progress, we explore the case with a thin coating, where the methods are more subtle and some preparation of the samples is needed. We
will address this issue as well as more advanced morphological parameters in a separate
publication.
 
\section*{Acknowledgements}
Most of the computations described in this paper have been carried out at the cluster of the Centre of Informatics Resources of Higher Normandy (CRIHAN - Centre de Ressources Informatiques de Haute-Normandie).\\
This work has been supported by ACCEA project selected by the "Fonds Unique Interministériel (FUI) 15 (18/03/2013)" program.




\newpage
\begin{thebibliography}{9}
\bibitem{Salnikov1}
 V. Salnikov, D. Choi, Ph. Karamian-Surville
On efficient and reliable stochastic generation of RVEs for analysis of composites within the framework of homogenization, 
Computational Mechanics, Volume 55, Issue 1, 2015. Preprint: \url{http://arxiv.org/abs/1408.6074}, 2014

\bibitem{CSMA}
 S. Lemaitre, V. Salnikov, D. Choi, P. Karamian
Approche par la dynamique moléculaire pour la conception de VER 3D et variations autour de la pixellisation, accepted for publication in the proceedings of the CSMA 2015 \url{https://csma2015.csma.fr/})

\bibitem{Eyre and Milton}
 D.J. Eyre, G.W. Milton 
A fast numerical scheme for computing the response of composites using grid refinement, Journal of Physique III 1999; 6; 41-47

\bibitem{Salnikov2}
 V. Salnikov, S. Lemaitre, D. Choi, Ph. Karamian-Surville
Measure of combined effects of morphological parameters of inclusions within composite materials via stochastic homogenization to determine effective mechanical properties, 
To appear in Composite Structures. DOI:10.1016/j.compstruct.2015.03.076, 2015, Preprint:
\url{http://arxiv.org/abs/1411.4037}


\bibitem{Moulinec and Silva}
 H. Moulinec, F. Silva
Comparison of the three accelerated FFT-based schemes for computing the mechanical response of composite materials, Int. J. Numer. Meth. Engng. 2014; 97:960-985

\bibitem{Sanchez}
 E. Sanchez-Palencia
Non-homogeneous media and vibration theory, vol. 127, Berlin, 1980)

\bibitem{Bensoussan}
A. Bensoussan, J.L. Lions, G.C. Papanicolaou
Asymptotic Analysis for Periodic Structures, Springer Verlag, North Hollland, Amsterdam Edition, 1978)

\bibitem{Moulinec and Sucquet 2001}
 J.C. Michel,H. Moulinec,  P. Suquet
A computational scheme for linar and non-linear composites with arbitrary phase contrast , Int. J. Numer. Meth. Engng. 2001; 52:139-160(DOI: 10.1002/nme.275)
\end{thebibliography}
\end{document}